\documentstyle[12pt]{article}
\def \b{{\cal B}}
\def \beq{\begin{equation}}
\def \eeq{\end{equation}}
\begin{document}
\rightline{CERN-TH/97-185}
\rightline{EFI 97-34}
\rightline{July 1997}
\rightline{hep-ph/9707521}
\bigskip

\centerline{\bf $B$ DECAYS INVOLVING $\eta$ AND $\eta'$}
\centerline{\bf IN THE LIGHT OF THE $B \to K \eta'$ PROCESS}
\bigskip
\centerline{Amol S. Dighe}
\centerline{\it Enrico Fermi Institute and Department of Physics,
University of Chicago}
\centerline{\it 5640 S. Ellis Avenue, Chicago IL 60615}
\smallskip
\centerline{and}
\smallskip
\centerline{Michael Gronau\footnote{Permanent Address: Physics
Department,
Technion -- Israel Institute of Technology, 32000 Haifa, Israel}}
\centerline{\it Theoretical Physics Division, CERN}
\centerline{\it CH-1211 Geneva 23, Switzerland}
\smallskip
\centerline{and}
\smallskip
\centerline{Jonathan L. Rosner}
\centerline{\it Enrico Fermi Institute and Department of Physics,
University of Chicago}
\centerline{\it 5640 S. Ellis Avenue, Chicago IL 60615}
\bigskip

\centerline{\bf ABSTRACT}
\bigskip

\begin{quote}
The observation by the CLEO Collaboration of the decays $B^{(+,0)} \to
K^{(+,0)} \eta'$ is shown to imply a significant but still uncertain
contribution from the flavor-SU(3)-singlet component of the $\eta'$.  By
comparing the rate for these decays with others for decays of $B$ mesons to two
pseudoscalar mesons, it is shown that the prospects for observing CP-violating
asymmetries in certain modes such as $B^+ \to \pi^+ \eta$ and $B^+ \to \pi^+
\eta'$ are quite bright. 
\end{quote}
\bigskip

\leftline{PACS numbers: 13.25.Hw, 14.40.Nd, 11.30.Er, 12.15.Ji}
\vfill
\leftline{CERN-TH/97-185}
\leftline{July 1997}
\newpage

The CLEO Collaboration \cite{CLEO} has recently reported the decays $B^+ \to
K^+ \eta'$ and $B^0 \to K^0 \eta'$ with branching ratios of $(7.1^{+2.5}_{-2.1}
\pm 0.9) \times 10^{-5}$ and $(5.3^{+2.8}_{-2.2} \pm 1.2) \times 10^{-5}$,
respectively. In the present note we show that these results, when combined
with other information on decays of $B$ mesons to pairs of light pseudoscalar
mesons, indicate that the $K \eta'$ decays receive a significant contribution
from the flavor-SU(3)-singlet component of the $\eta'$.  We use present
information to predict the rates for charged and neutral $B$'s to decay to
($\pi$ or $K) + (\eta$ or $\eta'$).  By searching for processes in which
contributions of different weak decay amplitudes are comparable to one another,
we show that there is a high likelihood for observable CP-violating asymmetries
in the decays $B^+ \to \pi^+ \eta$ and $B^+ \to \pi^+ \eta'$. A similar
conclusion was reached earlier by Barshay, Rein, and Sehgal \cite{BRS} on the
basis of a different analysis.  Others have emphasized previously the potential
for CP-violating rate asymmetries to be exhibited in decays of $B$ mesons to
pairs of charmless mesons \cite{Asymms}. 

The contribution to $B^+ \to K^+ \eta'$ from a new penguin amplitude, occurring
only in decays involving a flavor SU(3) singlet component in the final
pseudoscalar meson state, was noted in Refs.~\cite{Quad,DGRPL,ASD}. While one
possibility for this contribution \cite{BRS,KB,int} is an intrinsic $c \bar c$
component in the $\eta'$, more conventional mechanisms \cite{int,other} (e.g.,
involving gluons) also seem adequate to explain the observed rate.  Some
enhancement of conventional mechanisms may be needed to explain the large rate
for the {\em inclusive} process $B^+ \to \eta' + X$. We shall not be concerned
here with the inclusive process. 

We list the relevant decay amplitudes associated with a flavor-SU(3)
decomposition \cite{DGRPL,ASD,Zepp,SW,Chau,GHLR,GHLRS,GHLRP} in Tables I and
II. Unprimed amplitudes denote $\Delta S = 0$ decays; primed amplitudes denote
$|\Delta S| = 1$ decays.  An amplitude $t~(t')$ describes a tree-graph
contribution, $c~(c')$ describes a color-suppressed process, $p~(p')$ a penguin
graph contribution coupling to a pair of quark-antiquark mesons, and $s~(s')$
an additional penguin contribution coupling specifically to the
flavor-SU(3)-singlet component of the $\eta$ or $\eta'$.  All these amplitudes
are defined in such a way \cite{GHLRP} as to include contributions from
electroweak penguin terms \cite{EWP}. 

\begin{table}
\caption{Summary of predicted contributions to selected $\Delta S = 0$ decays
of $B$ mesons.  Rates are quoted in branching ratio units of $10^{-6}$.  Rates
in italics are assumed inputs.} 
\begin{center}
\begin{tabular}{r c c c c c c} \hline
Decay & Amplitudes & Denom. & $|t|^2$ & $|p|^2$ & \multicolumn{2}{c}
{$|s|^2$ rate} \\
      &            & factor &   rate  &   rate  &   (a) & (b) \\ \hline
$B^+ \to \pi^+ \pi^0$ & $t+c$ & $-\sqrt{2}$ & {\it 4.1} & 0 & 0 & 0 \\
$ \to K^+ \bar{K}^0$ & $p$ & 1 & 0 & 0.8 & 0 & 0 \\
$ \to \pi^+ \eta$ & $t+c+2p+s$ & $-\sqrt{3}$ & 2.8 & 1.0 & 0.06 & 0.24 \\
$ \to \pi^+ \eta'$ & $t+c+2p+4s$ & $\sqrt{6}$ & 1.4 & 0.5 & 0.4 & 1.9 \\
\hline
$B^0 \to \pi^+ \pi^-$ & $t+p$ & $-1$ & {\it 8.3} & 0.8 & 0 & 0 \\
$ \to \pi^0 \pi^0$ & $p-c$ & $\sqrt{2}$ & 0 & 0.4 & 0 & 0 \\
$ \to K^0 \bar{K}^0$ & $p$ & 1 & 0 & 0.8 & 0 & 0  \\
$ \to \pi^0 \eta$ & $2p+s$ & $-\sqrt{6}$ & 0 & 0.5 & 0.03 & 0.12 \\
$ \to \pi^0 \eta'$ & $p+2s$ & $\sqrt{3}$ & 0 & 0.26 & 0.2 & 0.9 \\ \hline
\end{tabular}
\end{center}

\leftline{(a): Constructive interference between $p'$ and $s'$ amplitudes
assumed in $B^+ \to K^+ \eta'$.}
\leftline{(b): No interference between $p'$ and $s'$ amplitudes
assumed in $B^+ \to K^+ \eta'$.}

\end{table}

\begin{table}
\caption{Summary of predicted contributions to selected $|\Delta S| = 1$ decays
of $B$ mesons.  Rates are quoted in branching ratio units of $10^{-6}$. Rates
in italics are assumed inputs.} 
\begin{center}
\begin{tabular}{r c c c c c c} \hline
Decay & Amplitudes & Denom. & $|t'|^2$ & $|p'|^2$ & \multicolumn{2}{c}
{$|s'|^2$ rate} \\
      &            & factor &   rate  &   rate  &  (a) & (b) \\ \hline
$B^+ \to K^0 \pi^+$ & $p'$ & 1 & 0 & {\it 16} & 0 & 0 \\
$ \to K^+ \pi^0$ & $t'+c'+p'$ & $-\sqrt{2}$ & 0.20 & 8 & 0 & 0 \\
$ \to K^+ \eta$ & $t'+c'+s'$ & $-\sqrt{3}$ & 0.13 & $\simeq 0$ & 1.2 & 4.9 \\
$ \to K^+ \eta'$ & $t'+c'+3p'+4s'$ & $\sqrt{6}$ & 0.07 & 24 & 9 & 39 \\
\hline
$B^0 \to K^+ \pi^-$ & $t'+p'$ & $-1$ & 0.4 & {\it 16} & 0 & 0 \\
$ \to K^0 \pi^0$ & $p'-c'$ & $\sqrt{2}$ & 0 & 8 & 0 & 0 \\
$ \to K^0 \eta$ & $c'+s'$ & $-\sqrt{3}$ & 0 & $\simeq 0$ & 1.2 & 4.9 \\
$ \to K^0 \eta'$ & $c'+3p'+4s'$ & $\sqrt{6}$ & 0 & 24 & 9 & 39 \\ \hline
\end{tabular}
\end{center}

\leftline{(a): Constructive interference between $p'$ and $s'$ amplitudes
assumed in $B^+ \to K^+ \eta'$.}
\leftline{(b): No interference between $p'$ and $s'$ amplitudes
assumed in $B^+ \to K^+ \eta'$.}

\end{table}

We assume the $\eta$ and $\eta'$ are mixed so that $\eta = ( u \bar u + d \bar
d - s \bar s)/\sqrt{3}$ and $\eta' = (u \bar u + d \bar d + 2 s \bar
s)/\sqrt{6}$, corresponding to an octet-singlet mixing angle of $\theta =
-19.5^{\circ}$.  The $p'$ contribution to $B \to K \eta$ vanishes for this
mixing \cite{GHLRS,GHLRP,HJL}.  More details justifying this assumption are
discussed, for example, in Refs.~\cite{Quad}, \cite{DGRPL}, and \cite{Chau}.
Other phase conventions for pseudoscalar mesons may be found in
Ref.~\cite{GHLR}.  We have neglected all annihilation- and exchange-type
amplitudes, which are expected to be highly suppressed in comparison with those
shown. 

In Tables I and II we have also calculated expected squares of contributions of
individual amplitudes to decays.  We ignore for present purposes any
interference between tree ($t$ or $t'$) and other amplitudes. We consider two
possibilities for the relative phase of the two predominant amplitudes, $p'$
and $s'$, in the decay $B^+ \to K^+ \eta'$.  The cases (a) and (b) listed in
the Tables correspond to constructive interference and no interference between
these amplitudes.  [Destructive interference would imply a singlet amplitude
$s'$ so large that the predicted value of $\b(B^+ \to K^+ \eta)$ would exceed
the current 90\% confidence level (c.l.) bound \cite{CLEO} $\b(B^+ \to K^+
\eta) < 8 \times 10^{-6}$.] 

Interference between amplitudes becomes important when they are not too
different in magnitude, which occurs in several cases which we shall identify
presently.  We do not quote contributions of color-suppressed amplitudes,
neglecting them in the ensuing discussion.  We determine amplitudes in the
following manner. 

(1) The magnitude of the $p'$ amplitude is estimated by averaging the observed
branching ratios \cite{Alex} 
\beq
\b(B^0 \to K^+ \pi^-) = (15^{+5+1}_{-4-1}
\pm 1) \times 10^{-6}
\eeq
and
\beq
\b(B^+ \to K^0 \pi^+) = (23^{+11+2}_{-10-2}
\pm 2) \times 10^{-6}
\eeq
to obtain the estimate $|p'|^2 = 16.3 \pm 4.3$, where all squares of amplitudes
in the Tables are quoted in branching ratio units of $10^{-6}$. In $B^0
\to K^+ \pi^-$ we have neglected the small $t'$ contribution, an
assumption which will be seen to be justified.  If the rates for (1) and (2)
are found to be unequal, the neglect of the $t'$ amplitude (or of some other
contribution) may not be valid.  In that case the possibility of a CP asymmetry
in (say) $B^0 \to K^+ \pi^-$ may be significantly enhanced. 

(2) The magnitude of the $p$ amplitude is estimated to be $|p| =
|V_{td}/V_{ts}| |p'|$, where $V_{td}$ and $V_{ts}$ are elements of the
Cabibbo-Kobayashi-Maskawa (CKM) matrix.  With an uncertainty of about a factor
of two, $|p|^2 \simeq |p'|^2/20 \simeq 0.8$.

(3) The $|p'|^2$ contribution to the decay $B^+ \to K^+ \pi^0$ is estimated to
be about 8, whereas \cite{Wurt}
\beq
\b(B^+ \to K^+ \pi^0) + \b(B^+ \to \pi^+ \pi^0) = (16^{+6+3}_{-5-2}
\pm 1) \times 10^{-6}~~~.
\eeq
Thus there is room for a significant $B^+ \to \pi^+ \pi^0$ signal. While 90\%
c.l. upper limits of $\b(B^+ \to \pi^+ \pi^0) < 20 \times 10^{-6}$ and
$\b(B^0 \to \pi^+ \pi^-) < 15 \times 10^{-6}$ are quoted in
Refs.~\cite{CLEO}, Ref.~\cite{Wurt} also quotes a $2.8 \sigma$ signal of 
\beq \label{eqn:ppz}
\b(B^+ \to \pi^+ \pi^0) = (9^{+6}_{-5}) \times 10^{-6}
\eeq
and a $2.2 \sigma$ signal of
\beq \label{eqn:pp}
\b(B^0 \to \pi^+ \pi^-) = (7 \pm 4) \times 10^{-6}~~~.
\eeq
Taking (\ref{eqn:ppz}) as an estimate of $|t|^2/2 = 9 \pm 5.5$ (neglecting the
color-suppressed amplitude $c$ in $B^+ \to \pi^+ \pi^0$), and (\ref{eqn:pp}) as
an estimate of $|t|^2 = 7 \pm 4$ (neglecting the penguin amplitude $p$ in
$B^0 \to \pi^+ \pi^-$), we find $|t|^2 = 8.3 \pm 3.8$. 

(4) The value of $|t'| = |V_{us}/V_{ud}| |t|$ is estimated without accounting
for SU(3)-breaking to lead to $|t'|^2 \simeq |t|^2/20 \simeq 0.4$.  It could be
slightly higher if one applied a correction \cite{GHLRS} of a factor of
$|f_K/f_\pi|^2$. 

(5) The $|p'|^2$ contribution to the $B \to K \eta'$ branching ratio (in units
of $10^{-6}$) is $(3/2)|p'|^2 = 24 \pm 6$; it cannot account for the observed
value of $63^{+20}_{-18}$ (our average for charged and neutral modes, where
$|t'|^2$ and $|c'|^2$ contributions are assumed to be negligible). Assuming
constructive interference between $p'$ and $s'$ in $B \to K \eta'$ we find the
$|s'|^2$ contribution to the rate to be about $(8/3)|s'|^2 = 9$, with an
additional contribution of 30 from the $s'$--$p'$ interference term.  [The
enhancement of the $B \to K \eta'$ rate by a modest $s'$ amplitude interfering
constructively with $p'$ was noted by Lipkin, last of Refs.~\cite{HJL}.] If the
interference term is absent (i.e., if the relative phase of the amplitudes is
$\pi/2$) then one needs an $|s'|^2$ contribution of $(8/3)|s'|^2 = 39$ to the
rate.  Henceforth we shall work only with central values of amplitudes for
illustrative purposes; the uncertainty in $|s'|$ due to the uncertainty in its
phase relative to $|p'|$ generally exceeds that due to experimental error.  (If
one allows the $B \to K \eta'$ branching ratio to be at its $-1 \sigma$ value,
$s'$ can even be considerably smaller, with $(8/3)|s'|^2 \simeq 4$, when $s'$
and $p'$ interfere constructively in this decay.) 

(6) Since we expect $|s/s'| = |V_{td}/V_{ts}|$ if both $s$ and $s'$ are
dominated by the top quark, we choose $|s|^2 = |s'|^2/20$. (If in fact $|s/s'|
= |V_{cd}/V_{cs}|$ as a result of charmed quark dominance of this type of
penguin contribution, the result is the same.) 

The results in the Tables may be interpreted in the following manner. 

(i) Any contribution of order 10 or greater (corresponding to a branching ratio
of $10^{-5}$) has been observed.

(ii) A contribution greater or equal than 1 should be observable in the next
generation of CLEO experiments, with improved sensitivity and particle
identification.  Thus the decays $B^0 \to \pi^+ \pi^-$, $B^0 \to K^0 \pi^0$,
$B^+ \to K^+ \eta$, $B^+ \to \pi^+ \eta$, and $B^+ \to \pi^+ \eta'$ should all
make their appearance, while $B^+ \to K^+ \pi^0$ and $B^+ \to \pi^+ \pi^0$
should be resolved from one another.  For example, one expects $\b(B^+ \to
\pi^+ \eta)\simeq 4 \times 10^{-6}$ and $\b(B^+ \to \pi^+ \eta') = (2~{\rm
to}~4) \times 10^{-6}$, where the current upper bounds \cite{CLEO} are $8
\times 10^{-6}$ and $45 \times 10^{-6}$, respectively. The first limit is
already quite close to our prediction. The above branching ratios are about a
factor of 2 larger than those predicted in Ref.~\cite{BRS}. 

(iii) The amplitudes for $B^0 \to K^0 \pi^0$, $B^0 \to K^0 \eta$, and $B^0 \to
K^0 \eta'$ satisfy
\beq
3 \sqrt{2} A(B^0 \to K^0 \pi^0) - 4 \sqrt{3} A(B^0 \to K^0 \eta) =
\sqrt{6}  A(B^0 \to K^0 \eta')~~~.
\eeq 
Aside from small $c'$ contributions, the terms in this triangle relation are
dominated by $p'$ and $s'$ contributions.  Since $p'$ and $s'$ are expected to
have the same weak phase, the shape of the triangle will tell us about the
relative strong phase of these amplitudes.  Neglecting $t'$ contributions as
well, one can write 
\beq
3 A(B^+ \to K^0 \pi^+) - 4 \sqrt{3} A(B^+ \to K^+ \eta) =
\sqrt{6}  A(B^+ \to K^+ \eta')~~~,
\eeq 
which is easier to measure.  The main uncertainty lies in the value of the
branching ratio for $B^+ \to K^+ \eta$.  If $s'$ involves strong rescattering
from charm-anticharm states \cite{BRS,Asymms,int,other}, its strong phase could
differ from that of $p'$ (and hence also possibly $t'$). 

(iv) Processes with two contributions both of which exceed 1 are prime
candidates for observable direct CP violation if both strong and weak phases of
the two amplitudes differ from one another.  The weak phases of the $t$ (tree)
and $p$ (penguin) contributions in $B^+ \to \pi^+ \eta$ are expected to be
$\gamma$ and $-\beta$ \cite{BF}, respectively, while the relative strong phases
are unknown.  In the case of $B^+ \to \pi^+ \eta'$, if its $s$ contribution is
dominated by the charmed quark penguin, a significant strong phase shift could
arise from the real $c \bar c$ intermediate state. One could thus have a large
strong phase shift difference between the $s$ and $t$ amplitudes in $B^+ \to
\pi^+ \eta'$. The weak phases of these two amplitudes are also different:  the
charm penguin is approximately real, while the $t$ amplitude has a weak phase
$\gamma$. 

(v) Our focus has been on the observability of direct CP violation in decays
such as $B^+ \to \pi^+ \eta$ and $B^+ \to \pi^+ \eta'$.  These processes may
not be the first to exhibit CP asymmmetries;  asymmetric $B$ factories will
search for mixing-induced asymmetries, in which the time dependence of the
decays must be studied. A time-dependent asymmetry measurement in $B^0\to
K_S\eta'$ would provide a clean determination of the weak phase $\beta$. Our
result, $|p/t| = 0.3$, implies a rather large ``penguin pollution'' in the
analysis of the time-dependent decay asymmetry in $B^0 \to \pi^+ \pi^-$,
compatible with previous estimates \cite{GHLRS,SilWo}. In order to resolve such
effects using isospin symmetry \cite{GL}, one would have to measure $B^0 \to
\pi^0 \pi^0$, for which the $|p|^2$ contribution to the branching ratio is only
$0.4 \times 10^{-6}$.  (The contribution of the color-suppressed amplitude $c$
is highly uncertain but unlikely to be much larger.) An alternative way to
resolve the penguin pollution question in $B^0 \to \pi^+ \pi^-$ is to rely on
flavor SU(3) to link this decay with various $B \to K \pi$ modes
\cite{SilWo,DGR}, all of which have large rates. 

To summarize, we have used existing data on $B \to K \eta'$ and other two-body
modes involving pairs of light pseudoscalar mesons to anticipate observable
CP-violating effects in the decays $B^+ \to \pi^+ \eta$ and $B^+ \to \pi^+
\eta'$.  Since experimental errors are still quite large, the same procedure,
based only on flavor SU(3), can and should be applied to better data when they
become available.  In that case one will be able to test for effects of
interference among various amplitudes which have been ignored here (applying,
for example, amplitude relations noted in Refs.~\cite{Quad,DGRPL,ASD}).  One
welcome improvement in data will be a better estimate of $|p/p'|^2$ and
$|s/s'|^2$, which, under the assumption of top quark dominance, we have taken
to be $|V_{td}/V_{ts}|^2 \simeq 1/20$, with an uncertainty of a factor of 2. 

A rule of thumb for observable CP-violating effects is that one must at least
be able to observe the {\it square} of the lesser of two interfering amplitudes
at the $n \sigma$ level in order to observe an asymmetry at this level
\cite{DR}.  Our results indicate that this sensitivity threshold is passed for
decays of the form $B^+ \to \pi^+ \eta$ and $B^+ \to \pi^+ \eta'$ when
branching ratios of order $10^{-6}$ become detectable in experiments sensitive
to both charged and neutral final-state particles. 

We are grateful to J. Alexander, D. Atwood, K. Berkelman, P. Drell, H. Fritzsch
and L. Sehgal for helpful discussions. This work was performed in part at the
Aspen Center for Physics, and supported in part by the United States Department
of Energy under Grant No.~DE FG02 90ER40560 and by the United States -- Israel
Binational Science Foundation under Research Grant Agreement 94-00253/2. 
\bigskip

\def \ajp#1#2#3{Am. J. Phys. {\bf#1}, #2 (#3)}
\def \apny#1#2#3{Ann. Phys. (N.Y.) {\bf#1}, #2 (#3)}
\def \app#1#2#3{Acta Phys. Polonica {\bf#1}, #2 (#3)}
\def \arnps#1#2#3{Ann. Rev. Nucl. Part. Sci. {\bf#1}, #2 (#3)}
\def \art{and references therein}
\def \cmts#1#2#3{Comments on Nucl. Part. Phys. {\bf#1}, #2 (#3)}
\def \cn{Collaboration}
\def \cp89{{\it CP Violation,} edited by C. Jarlskog (World Scientific,
Singapore, 1989)}
\def \dpfa{{\it The Albuquerque Meeting: DPF 94} (Division of Particles and
Fields Meeting, American Physical Society, Albuquerque, NM, Aug.~2--6, 1994),
ed. by S. Seidel (World Scientific, River Edge, NJ, 1995)}
\def \dpff{{\it The Fermilab Meeting: DPF 92} (Division of Particles and Fields
Meeting, American Physical Society, Batavia, IL., Nov.~11--14, 1992), ed. by
C. H. Albright \ite~(World Scientific, Singapore, 1993)}
\def \efi{Enrico Fermi Institute Report No. EFI}
\def \epl#1#2#3{Europhys.~Lett.~{\bf #1}, #2 (#3)}
\def \f79{{\it Proceedings of the 1979 International Symposium on Lepton and
Photon Interactions at High Energies,} Fermilab, August 23-29, 1979, ed. by
T. B. W. Kirk and H. D. I. Abarbanel (Fermi National Accelerator Laboratory,
Batavia, IL, 1979}
\def \hb87{{\it Proceeding of the 1987 International Symposium on Lepton and
Photon Interactions at High Energies,} Hamburg, 1987, ed. by W. Bartel
and R. R\"uckl (Nucl. Phys. B, Proc. Suppl., vol. 3) (North-Holland,
Amsterdam, 1988)}
\def \ib{{\it ibid.}~}
\def \ibj#1#2#3{~{\bf#1}, #2 (#3)}
\def \ichep72{{\it Proceedings of the XVI International Conference on High
Energy Physics}, Chicago and Batavia, Illinois, Sept. 6 -- 13, 1972,
edited by J. D. Jackson, A. Roberts, and R. Donaldson (Fermilab, Batavia,
IL, 1972)}
\def \ijmpa#1#2#3{Int. J. Mod. Phys. A {\bf#1}, #2 (#3)}
\def \ite{{\it et al.}}
\def \jpb#1#2#3{J.~Phys.~B~{\bf#1}, #2 (#3)}
\def \lkl87{{\it Selected Topics in Electroweak Interactions} (Proceedings of
the Second Lake Louise Institute on New Frontiers in Particle Physics, 15 --
21 February, 1987), edited by J. M. Cameron \ite~(World Scientific, Singapore,
1987)}
\def \ky85{{\it Proceedings of the International Symposium on Lepton and
Photon Interactions at High Energy,} Kyoto, Aug.~19-24, 1985, edited by M.
Konuma and K. Takahashi (Kyoto Univ., Kyoto, 1985)}
\def \mpla#1#2#3{Mod. Phys. Lett. A {\bf#1}, #2 (#3)}
\def \nc#1#2#3{Nuovo Cim. {\bf#1}, #2 (#3)}
\def \np#1#2#3{Nucl. Phys. {\bf#1}, #2 (#3)}
\def \pisma#1#2#3#4{Pis'ma Zh. Eksp. Teor. Fiz. {\bf#1}, #2 (#3) [JETP Lett.
{\bf#1}, #4 (#3)]}
\def \pl#1#2#3{Phys. Lett. {\bf#1}, #2 (#3)}
\def \pla#1#2#3{Phys. Lett. A {\bf#1}, #2 (#3)}
\def \plb#1#2#3{Phys. Lett. B {\bf#1}, #2 (#3)}
\def \pr#1#2#3{Phys. Rev. {\bf#1}, #2 (#3)}
\def \prc#1#2#3{Phys. Rev. C {\bf#1}, #2 (#3)}
\def \prd#1#2#3{Phys. Rev. D {\bf#1}, #2 (#3)}
\def \prl#1#2#3{Phys. Rev. Lett. {\bf#1}, #2 (#3)}
\def \prp#1#2#3{Phys. Rep. {\bf#1}, #2 (#3)}
\def \ptp#1#2#3{Prog. Theor. Phys. {\bf#1}, #2 (#3)}
\def \ptwaw{Plenary talk, XXVIII International Conference on High Energy
Physics, Warsaw, July 25--31, 1996}
\def \rmp#1#2#3{Rev. Mod. Phys. {\bf#1}, #2 (#3)}
\def \rp#1{~~~~~\ldots\ldots{\rm rp~}{#1}~~~~~}
\def \si90{25th International Conference on High Energy Physics, Singapore,
Aug. 2-8, 1990}
\def \slc87{{\it Proceedings of the Salt Lake City Meeting} (Division of
Particles and Fields, American Physical Society, Salt Lake City, Utah, 1987),
ed. by C. DeTar and J. S. Ball (World Scientific, Singapore, 1987)}
\def \slac89{{\it Proceedings of the XIVth International Symposium on
Lepton and Photon Interactions,} Stanford, California, 1989, edited by M.
Riordan (World Scientific, Singapore, 1990)}
\def \smass82{{\it Proceedings of the 1982 DPF Summer Study on Elementary
Particle Physics and Future Facilities}, Snowmass, Colorado, edited by R.
Donaldson, R. Gustafson, and F. Paige (World Scientific, Singapore, 1982)}
\def \smass90{{\it Research Directions for the Decade} (Proceedings of the
1990 Summer Study on High Energy Physics, June 25--July 13, Snowmass, Colorado),
edited by E. L. Berger (World Scientific, Singapore, 1992)}
\def \tasi90{{\it Testing the Standard Model} (Proceedings of the 1990
Theoretical Advanced Study Institute in Elementary Particle Physics, Boulder,
Colorado, 3--27 June, 1990), edited by M. Cveti\v{c} and P. Langacker
(World Scientific, Singapore, 1991)}
\def \waw{XXVIII International Conference on High Energy
Physics, Warsaw, July 25--31, 1996}
\def \yaf#1#2#3#4{Yad. Fiz. {\bf#1}, #2 (#3) [Sov. J. Nucl. Phys. {\bf #1},
#4 (#3)]}
\def \zhetf#1#2#3#4#5#6{Zh. Eksp. Teor. Fiz. {\bf #1}, #2 (#3) [Sov. Phys. -
JETP {\bf #4}, #5 (#6)]}
\def \zpc#1#2#3{Zeit. Phys. C {\bf#1}, #2 (#3)}
\def \zpd#1#2#3{Zeit. Phys. D {\bf#1}, #2 (#3)}


\begin{thebibliography}{99}

\bibitem{CLEO} CLEO Collaboration, Cornell University report CLEO CONF 97-22,
presented to Lepton--Photon Symposium, Hamburg, July 1997.  See also J. Smith
at the 1997 Aspen Winter Physics Conference on Particle Physics, Aspen, CO,
January, 1997; B. Behrens and J. Alexander at {\it The Second International
Conference on $B$ Physics and CP Violation}, Honolulu, HI, March 24--27, 1997;
and F. W\"urthwein at {\it Les Rencontres du Moriond: QCD and High Energy
Hadronic Interactions}, Les Arcs, France, March 1997, hep-ex/9706010.  In
quoting rates we do not distinguish between a process and its charge-conjugate,
though such a difference is precisely what we search for when looking for
direct CP violation. 

\bibitem{BRS} S. Barshay, D. Rein, and L. M. Sehgal, \plb{259}{475}{1991}.

\bibitem{Asymms} M. Bander, D. Silverman, and A. Soni, \prl{43}{242}{1979}; J.
M. G\'erard and W. S. Hou, \prl{62}{855}{1989}; \prd{43}{2909}{1991};
G. Kramer, W. F. Palmer, and H. Simma, \np{B428}{77}{1994}; \zpc{66}{429}
{1995}.

\bibitem{Quad} M. Gronau and J. L. Rosner, \prd{53}{2516}{1996}.

\bibitem{DGRPL} A. S. Dighe, M. Gronau, and J. L. Rosner, \plb{367}{357}{1996};
\ibj{377}{325(E)}{1996}; 

\bibitem{ASD} A. S. Dighe, \prd{54}{2067}{1996}.

\bibitem{KB} K. Berkelman, CLEO note CBX 96-79 and supplement, unpublished.

\bibitem{int} I. Halperin and A. Zhitnitsky, hep-ph/9704412 and hep-ph/9705251; 
F. Yuan and K.-T. Chao, \prd{56}{R2495}{1997}; A. Ali and C. Greub, DESY
97-126, hep-ph/9707251. 

\bibitem{other} D. Atwood and A. Soni, \plb{405}{150}{1997}; hep-ph/9706512;
W.-S. Hou and B. Tseng, hep-ph/9705304; H.-Y. Cheng and B. Tseng,
IP-ASTP-03-97/NTU-TH-97-08, hep-ph/9707316; A. Datta, X.-G. He, and S. Pakvasa,
hep-ph/9707259; A. L. Kagan and A. A. Petrov, UCHEP-27/UMHEP-443,
hep-ph/9707354; H. Fritzsch, CERN-TH/97-200, hep-ph/9708348. 

\bibitem{Zepp} D. Zeppenfeld, \zpc{8}{77}{1981}.

\bibitem{SW} M. Savage and M. Wise, \prd{39}{3346}{1989};
\ibj{40}{3127(E)}{1989}.

\bibitem{Chau} L. L. Chau {\it et al.}, \prd{43}{2176}{1991}.

\bibitem{GHLR} M. Gronau, O. Hern\'andez, D. London, and J. L. Rosner,
\prd{50}{4529}{1994}.

\bibitem{GHLRS} M. Gronau, O. Hern\'andez, D. London, and J. L. Rosner,
\prd{52}{6356}{1995}.

\bibitem{GHLRP} M. Gronau, O. Hern\'andez, D. London, and J. L. Rosner,
\prd{52}{6374}{1995}.

\bibitem{EWP} R. Fleischer, \zpc{62}{81}{1994}; \plb{321}{259}{1994};
\ibj{332}{419}{1994}; N. G. Deshpande and X.-G. He, \plb{336}{471}{1994};
\prl{74}{26}{1995}; N. G. Deshpande, X.-G. He, and J. Trampetic,
\plb{345}{547}{1995}.

\bibitem{HJL} The suppression of the $p'$ contribution to the $K \eta$ mode and
its enhancement for the $K \eta'$ mode is a special case of a more general
mechanism noted by H. J. Lipkin, \prl{46}{1307}{1981}; \plb{254}{247}{1991},
\art; Argonne National Laboratory report ANL-HEP-CP-97-45, presented at {\it
The Second International Conference on $B$ Physics and CP Violation}, Honolulu,
HI, March 24--27, 1997. [See also Ref.~\cite{Chau}, where the relatively large
value $\b(B^+ \to K^+ \eta') = 3.6 \times 10^{-5}$ is predicted.]  Lipkin
points out that the same mechanism works in the opposite direction for $K^* +
(\eta,\eta')$ final states, enhancing $K^* \eta$ and suppressing $K^* \eta'$
decays. 

\bibitem{Alex} J. Alexander, Ref.~\cite{CLEO}.

\bibitem{Wurt} F. W\"urthwein, Ref.~\cite{CLEO}.

\bibitem{BF} The phase $- \beta~[= {\rm Arg}(V_{td})]$ in the $b \to d$ penguin
term may receive corrections from $u$ and $c$ penguins:  See A. J. Buras and R.
Fleischer, \plb{341}{379}{1995}. 

\bibitem{SilWo} J. Silva and L. Wolfenstein, \prd{49}{R1151}{1994}.

\bibitem{GL} M. Gronau and D. London, \prl{65}{3381}{1990}.

\bibitem{DGR} M. Gronau and J. L. Rosner, \prl{76}{1200}{1996}; A. S. Dighe, M.
Gronau, and J. L. Rosner, \prd{54}{3309}{1996}; A. S. Dighe and J. L. Rosner,
\prd{54}{4677}{1996}.

\bibitem{DR} I. Dunietz and J. L. Rosner, \prd{34}{1404}{1986}. 
\end{thebibliography}
\end{document}